# Unusual Red Light Emission from Nonmetallic Cu$_2$Te Microdisk for Laser and SERS Applications

*Qiuguo Li,* [1] *Hao Rao,* [2] *Xinzhou Ma,* [3] *Haijuan Mei,* [1] *Zhengting Zhao,* [1] *Weiping Gong,\*,* [1] *Andrea Camposeo* [4]*, Dario Pisignano* [4,5] *and Xianguang Yang\*,* [2]

[1]Guangdong Provincial Key Laboratory of Electronic Functional Materials and Devices, Huizhou University, Huizhou 516001, Guangdong, China

[2]Institute of Nanophotonics, Jinan University, Guangzhou 511443, China

[3]School of Materials Science and Energy Engineering, Foshan University, Foshan, 528000, China

[4]NEST, Istituto Nanoscienze-CNR and Scuola Normale Superiore, Piazza S. Silvestro 12, I-56127 Pisa, Italy

[5]Dipartimento di Fisica, Università di Pisa, Largo B. Pontecorvo 3, I-56127 Pisa, Italy

**ABSTRACT:**

Physical characteristics of $Cu_2Te$ are poorly investigated due to limited Te sources available and unclear atomic positions of crystal structure. Herein, hexagonal $Cu_2Te$ microdisks are successfully prepared via chemical vapor deposition procedure using GaTe as Te source. The epitaxial growth mechanism of the $Cu_2Te$ hexagonal structures with the orthorhombic phase are rationalized by proposed layer-over-layer growth model. The photoluminescence (PL) spectrum of $Cu_2Te$ microdisks shows a new red emission band in addition to usual infrared light emission due to Cu deficiency. Single $Cu_2Te$ microdisk operates as an optical microcavity supporting whispering gallery modes for red lasing around 627.5 nm. This $Cu_2Te$ microdisk microcavity exhibits a high quality factor of 1568 and a low lasing threshold of 125 kW·cm$^{-2}$ at room temperature. Meanwhile, $Cu_2Te$ microdisks have been exhibited as an ideal platform for surface enhanced Raman scattering (SERS) eliminating drawbacks of noble metal substrates with detection limitation to nanomolar level and an enhancement factor of $\sim 1.95 \times 10^5$. Hexagonal $Cu_2Te$ microdisks turn out to be an efficient microcavity for red lasing and low-cost nonmetallic SERS substrates, opening potential applications in photonics and biological detection of aromatic molecules.

1. **INTRODUCTION**

Copper telluride, usually p-type semiconductors due to copper vacancies and with bandgap ranging from 1.0 to 1.5 eV [1,2], can exist in several phases and compositions by tuning the ratio of Cu to Te. For the valuable electric, optical and thermal properties, such as localized surface plasmon resonance (LSPR) in the near infrared (NIR) spectral interval, copper telluride is investigated for solar cells[3], lithium ion batteries[4], surface enhanced Raman scattering (SERS) probe[5,6], thermoelectric[7,8] and photoelectrochemical applications[9,10]. Since 2010s, $Cu_{2-x}Te$ nanostructures with defined morphologies began to flourish. These nanostructures were prepared by solvothermal growth[11], microwave-assisted elemental reactions[12], and electrodeposition method[9]. Kriegel et al. introduced the first hot-injection synthesis of $Cu_{2-x}Te$ nanocubes, exhibiting pronounced LSPR properties in the NIR region[13]. Furthermore, $Cu_{2-x}Te$ nanostructures with well-defined morphologies (spheres, rods, and tetrapods) were obtained and used as sacrificial templates for the synthesis of CdTe nanostructures by a $Cu^+/Cd^{2+}$ cation exchange reaction[14]. In spite of the difficulty of balancing the reactivity and compatibility with other elemental precursors in the synthesis of $Cu_{2-x}Te$ nanostructures, $Cu_2Te$, that is one important compound of copper telluride class, till remain unclear and controversial. Two types of low-temperature crystalline phases for $Cu_2Te$ have been reported[15,16]: the hexagonal structure proposed by Nowotny[15] has been proved to be unstable, while an orthorhombic superstructure with much larger unit cell[16] has been proposed and determined experimentally, even though the detailed atomic positions for this superstructure remains unclear. Recently, first-principles density functional theory (DFT) calculations have shown that the superstructure system features a preferentially layered structure, with an associated lower total energy than those of the previous models[17]. Experimentally, two-dimensional (2D) atomic monolayer $Cu_2Te$ has been fabricated on a graphene-SiC (0001) substrate by molecular beam epitaxy (MBE) with chemical stability against air-exposure[18]. The phase sequence in $Cu_2Te$ is highly complicated for copper telluride because of five successive phase transitions between room temperature and 900 K[19-21]. Despite years of studies, the crystal structure for the binary $Cu_2Te$ compound has still not been well-determined[22-24]. Thus, the crystal structure determination is a prerequisite to unveil the atomic position in the $Cu_2Te$ structure and then circumvent the controversy. Furthermore, thermoelectric[25-29], electrocatalytic[30], plasmonics and photonics characteristics[6] of micro/nanoscale $Cu_2Te$ are poorly investigated due to reduced number of available tellurium sources. In this context, photothermal therapy[31-33] is emerging as a new application of $Cu_2Te$. Moreover, thin films of cuprous chalcogenides of $Cu_2Te$[34] have already been exploited in solar cell, whereas plasmonic and photonic properties of $Cu_2Te$ are basically limited to studies of the luminescence features. Therefore, it is highly desirable to fabricate $Cu_2Te$ nanostructures with controllable morphology and tunable bandgap for applications such as photoluminescence (PL), laser and SERS, since low-cost and abundant nonmetallic plasmonic materials have been regarded as a promising substitute of noble metals for laser and SERS applications.

In this paper, we report on the fabrication of high-quality $Cu_2Te$ microdisks on copper foam or sheet by chemical vapor deposition (CVD) using GaTe as Te source, thus eliminating the constraint of limited tellurium sources available. The crystal structure at atomic level and stoichiometry of the layered $Cu_2Te$ microdisks are determined by X-ray diffraction (XRD), transmission electron microscopy (TEM), selected area electron diffraction (SAED) and X-ray photoelectron spectroscopy (XPS) characterizations. The crystal structure of $Cu_2Te$ microdisk is demonstrated to be an orthorhombic superstructure. In addition, Raman characterization using 532 nm laser excitations is performed to further verify the atomic structure of $Cu_2Te$ microdisk.

Interestingly, PL characterization of $Cu_2Te$ microdisk shows an emission band centered around 627.5 nm with lasing features, which has never been observed before and might be due to the existence of Cu vacancies. Numerical simulations are performed to study the optical mode distribution in $Cu_2Te$ microdisk, evidencing strong whispering-gallery mode (WGM) in the microdisk cavity. Besides, $Cu_2Te$ microdisks, as low-cost and abundant reserved nonmetallic SERS substrates, further enhance charge transfer and exciton resonances in SERS detection of rhodamine 6G (R6G) molecules with a detection limit of $10^{-9}$ M and an enhancement factor of $\sim 1.95 \times 10^5$. The hexagonal $Cu_2Te$ micordisk are relevant for potential applications in light-emitting devices, microcavity lasers and SERS detection as a promising substitute of noble metals.

## 2. RESULTS AND DISCUSSION

For the morphology characterization, Figure 1a shows the scanning electron microscopy (SEM) image of the as-prepared $Cu_2Te$ microdisks grown on a copper foam, which has typical foam holes. A dense ensemble of $Cu_2Te$ microdisks is present on the surface of the copper foam. Figure 1b shows the magnified SEM image of an area selected from Figure 1a, which evidence $Cu_2Te$ microdisks with hexagonal shape, with side length of 2–5 µm and thickness of ~1 µm. Figure 1b clearly show the hexagonal structure, indicating that the structure of the grown $Cu_2Te$ microdisks is composed by stacked layers (see also Figure S1 of the Supporting Information). Namely, each $Cu_2Te$ microdisk is layer-over-layer stacked on the upper surface of adjacent microdisk with the same orientation, thus forming the stack-on-stack structure of $Cu_2Te$ microdisks. The smaller microdisks are stacked onto the larger one (forming clear terraces) either on the center or near the edge, depending on the growth condition and direction. These $Cu_2Te$ microdisks prefer to growth on the (2 1 1) and (3 4 6) planes, which are identified through XRD analysis (Fig. 2b). Moreover, the high-resolution TEM (HRTEM) experiments discussed below also provide strong evidence for the growth direction of $Cu_2Te$ microdisks. The preferred growth planes of microdisks are ascribed to the lattice planes for which the atoms feature a closest-packing and the minimum free energy. This stack-on-stack hexagonal structure of $Cu_2Te$ microdisks is highly beneficial for the optical excitation of the WGM supported in the microcavity, which contributes to the lasing behavior as discussed later.

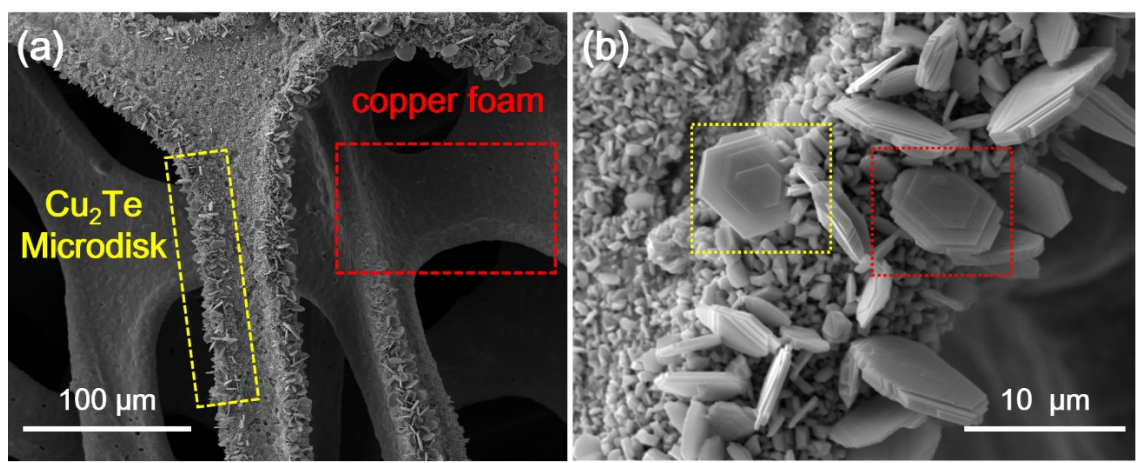

Figure 1. Morphology characterization. (a) SEM image of $Cu_2Te$ microdisks grown on a copper foam. (b) Magnified SEM image of a selected area from (a), showing the hexagonal structure.

X-ray spectroscopy (EDS) analysis is performed to obtain the composition of as-prepared $Cu_2Te$ microdisks. Cu, Te signals are observed in the EDS spectrum of the $Cu_2Te$ microdisks in Figure 2a, which confirms the elemental contents of Cu (atomic%, 65.68%) and Te (atomic%, 34.32%), giving the atomic ratio of Cu:Te to be 1.914:1. To confirm the crystal structure, XRD analysis is performed to investigate the crystallinity of the $Cu_2Te$ microdisks. Several diffraction peaks of $Cu_2Te$ microdisks in Figure 2b correspond to that of the orthorhombic phase $Cu_{0.664}Te_{0.336}$ (JCPDS PDF Card No. 00-037-1027). These peaks can be attributed to (0 3 1), (2 1 1), (2 2 2), (2 1 9), (0 9 0), (0 0 17), (3 4 6) and (0 1 18) lattice planes of $Cu_{0.664}Te_{0.336}$, while the peak (2 0 0) denoted as ★ is related to the copper foam. More specifically, the peaks at $2\theta = 12.23°$, $24.7°$, 42.11° and 43.25° correspond to the (0 3 1), (2 1 1), (0 0 17) and (3 4 6) planes, respectively. The Cu:Te atomic ratio of $Cu_{0.664}Te_{0.336}$ is about 1.976:1, which agrees with the ratio of 1.914:1 obtained from EDS analysis. The ratio difference may come from measurement errors. While the orthorhombic superstructure has been proposed and determined experimentally, the detailed atomic positions of this orthorhombic superstructure remain unsolved, with a lacking knowledge of structure, electronic and optical characteristics for $Cu_{0.664}Te_{0.336}$. First, the optical properties are characterized by Raman analysis. Room-temperature Raman spectra are acquired with same acquisition time (150 s) at excitation wavelengths of $\lambda_{ex}$ = 532 nm. Figure 2c shows the main Raman peaks of $Cu_{0.664}Te_{0.336}$ under 532 nm laser excitation, which are positioned at 75 and 117 cm$^{-1}$. These peaks are in agreement with the $B_{3g}$ and $B_1$ mode frequencies of the main components of the $Cu_2Te$ phase[35,36]. According to the Reference 36, the spectra of the samples from $Cu_{2-x}Te$ with ratios [Cu]/[Te] = 1.5 and 1.75 show an intense peak at 123 cm$^{-1}$ ($B_{2g}$ mode), while the one from $Cu_{2-x}Te$ with [Cu]/[Te] = 2 has a maximum located at 121 cm$^{-1}$. When 532 nm laser is focused onto the $Cu_{0.664}Te_{0.336}$ sample with a spot size of ~1μm, the Raman characteristics showed good stability with the pronounced $B_1$ mode peak during the laser intensity increased from 2.5 kW·cm$^{-2}$ to 125 kW·cm$^{-2}$. When light intensity was raised to 250 kW·cm$^{-2}$, the Raman peak shifted to 121 cm$^{-1}$, $B_{2g}$ mode become predominantly pronounced, maybe due to the temperature increase caused by high power laser and phase change at high temperature. Moreover, at high excitation intensity a broad Raman peak at 271 cm$^{-1}$ is observed and attributed to the Tellurium melting and then dissociation of the $Cu_{0.664}Te_{0.336}$ compounds[35]. Although at small wavenumbers (< 50–100 cm$^{-1}$) the acquired spectrum of $Cu_{0.664}Te_{0.336}$ compounds might overlap with the cutoff wavelength of the high frequency filters used for attenuating the Rayleigh scattering, a peak at 75 cm$^{-1}$ ($B_{3g}$ mode) can be distinguished especially at higher excitation intensity.

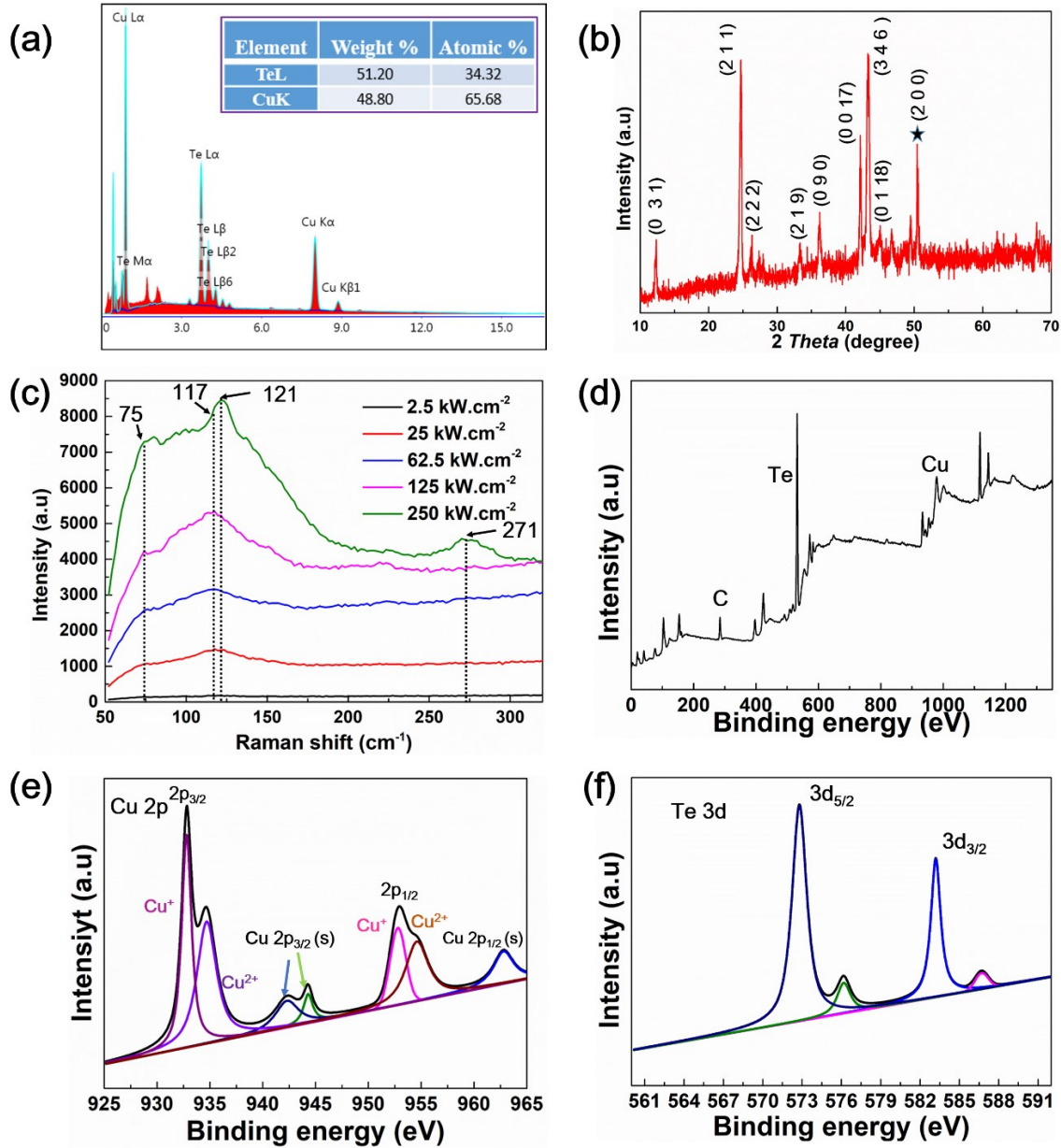

Figure 2. EDS, XRD, Raman and XPS characterizations. (a) EDS spectrum of the $Cu_2Te$ microdisks. Inset shows elemental contents. (b) XRD patterns of $Cu_2Te$ microdisks, the lattice plane of (2 0 0) for Cu are denoted as ★ near 50 degrees. (c) Raman spectra of the $Cu_2Te$ microdisks obtained under laser excitations ($\lambda_{ex}$ = 532 nm) with different light intensity at room temperature. (d, e, f) High resolution XPS spectra of (d) full spectrum and (e) Cu ($2p_{1/2}$ and $2p_{3/2}$) and (f) Te ($3d_{3/2}$ and $3d_{5/2}$) core levels.

To further investigate the surface valence state of the elements in the $Cu_2Te$, high-resolution XPS spectra of Cu and Te elements from $Cu_2Te$ microdisk sample are measured and shown in Figure 2e and 2f, respectively. Figure 2d shows the full XPS spectrum of $Cu_2Te$ with element denotation. In the XPS spectrum of Cu (Figure 2e), Cu $2p_{1/2}$ peaks centered at binding energy of 952.6 and 953.7 eV are present, while two peaks at 932.8 and 934.5 eV are attributed to Cu $2p_{3/2}$ in $Cu_2Te$, respectively[18]. In addition, two strong $Cu^{2+}$ peaks are observed maybe due to the vacancy of Cu in $Cu_2Te$. These results confirm the presence of divalent and univalent copper on the surface of $Cu_2Te$ microdisk, and the vacancy of Cu would modify the bandgap of $Cu_2Te$, which is relevant for bandgap engineering. Figure 2f shows two strong peaks of 584.0 and 573.4 eV, which are due to Te $3d_{3/2}$ and Te $3d_{5/2}$ in $Cu_2Te$, respectively[18]. The small peaks of 586.7 and 576.6 eV are possible indication of the presence of $TeO_2$, which is due to the oxidation of Te and/or the vacancy of Cu. The XPS results are in good agreement with our XRD data. Therefore, the XPS results suggest that the $Cu_2Te$ is successfully prepared and the Cu vacancy would be beneficial for bandgap engineering on optoelectronics and laser applications.

To obtain the detail crystal structure of $Cu_2Te$ microdisk, TEM and HRTEM measurements are performed. Figure 3a shows a typical TEM image of the $Cu_2Te$ microsheet, obtained by the sonication of $Cu_2Te$ microdisk into $Cu_2Te$ microsheet, supported on a nickel micro-grid. The hexagonal shape is destroyed during the sonication procedure. Figure 3b shows the magnified view of an area denoted with dashed line in Figure 3a. To further identify the crystallographic structure, Figure 3c shows the HRTEM image of an area denoted with dashed line in Figure 3b. The double spacing order exist within the (211) plane, which might be the contrast ratio of Moiré fringe obtained by TEM, specified in the two sets of crystal lattice with same period. The interplanar spacing of 3.592 Å can be attributed to the lattice plane of (2 1 1) in $Cu_2Te$ with orthorhombic phase. Another interplanar spacing of 2.097 Å can be identified from the lattice fringes in Figure 3d, corresponding to the lattice plane of (3 4 6) in $Cu_2Te$ with orthorhombic phase. The above findings rule out the occurrence of phase transformation in $Cu_2Te$. The morphology change can be the result of the mechanical modification of the microdisk shape. The selected area electron diffraction (SAED) pattern of the Cu$_2$Te microdisk in Figure 3e also confirms the crystal structure of Cu$_2$Te with orthorhombic phase, which is consistent with the XRD results.

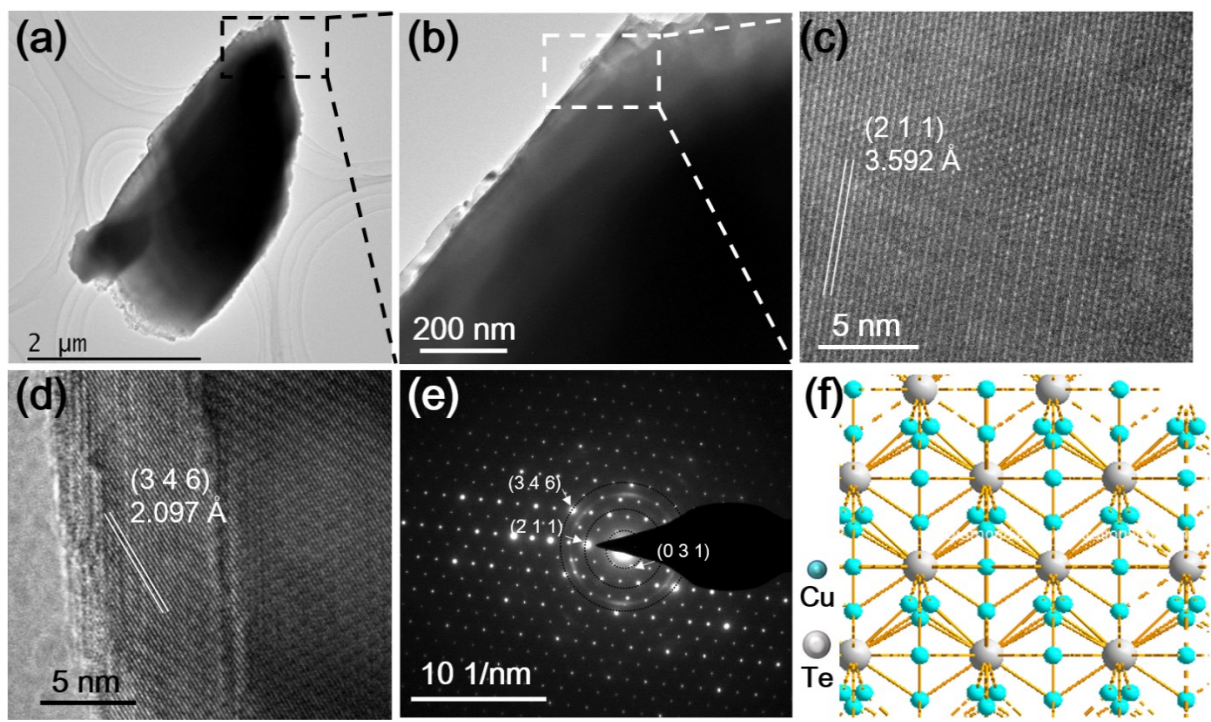

Figure 3. TEM, HRTEM and SAED characterizations. (a) Low-magnification and (b) high-magnification TEM images of single Cu$_2$Te microsheet. (c, d) HRTEM images of the Cu$_2$Te microdisk along the (2 1 1) and (3 4 6) directions with interplanar spacing of (c) 3.592 Å and (d) 2.097 Å. (e) SAED pattern of the Cu$_2$Te microsheet. (f) Schematic representation of the <0 1 0> side-view crystal lattices of the Cu$_2$Te with orthorhombic phase.

Schematically, Figure 3f shows the <0 1 0> side-view crystal lattices of the Cu$_2$Te with orthorhombic phase. All these characterizations demonstrate the high crystallinity of the as-synthesized Cu$_2$Te microdisks, which can be exploited for the light amplification in these naturally-formed whispering gallery cavities.

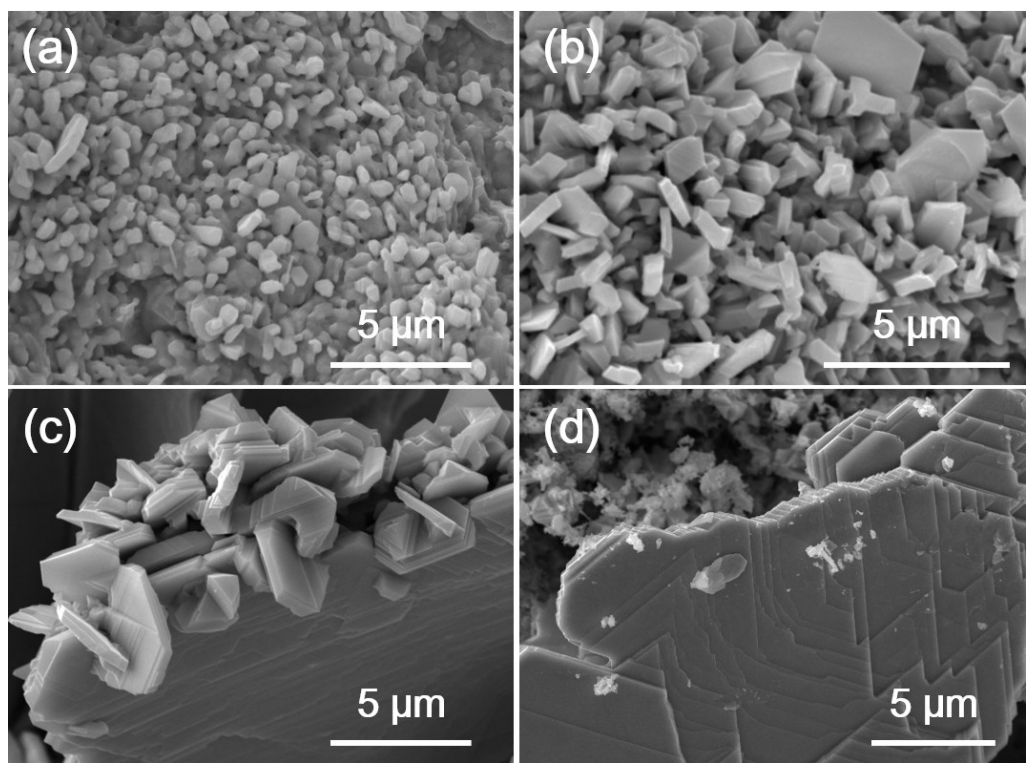

Figure 4. Surface morphology of $Cu_2Te$ microdisks at different growth times. Where the growth time for the CVD deposition time of GaTe is (a) 5 min, (b) 10 min, (c) 30 min and (d) 50 min, respectively.

To investigate the growth mechanism of $Cu_2Te$ microdisks, a series of experiments are performed to analyze the evolution of the surface morphology. Figure 4 shows the surface morphology of $Cu_2Te$ microdisks at different growth times of (a) 5 min, (b) 10 min, (c) 30 min, and (d) 50 min. For 5 min of the CVD deposition time of GaTe, small microdisks with size of 500 nm are observed on the copper foam, which may act as nucleation sites for $Cu_2Te$ growth. With deposition time increased up to 10 min, hexagonal type $Cu_2Te$ microdisks with large size up to 1 μm emerged (see Figure 4b), indicating the turning point for the formation of microdisks. The layer-over-layer morphology of $Cu_2Te$ microdisks is mainly observed on the basal surface of $Cu_2Te$ microdisks (see Figure 4c) as the deposition time increased up to 30 min. Once the deposition time reached 50 min, the $Cu_2Te$ microdisks are contiguously covering the copper foam with triangle shape and stack-on-stack structure (see Figure 4d). Based on the above results on experimental morphology, a layer-over-layer growth model for stack-on-stack deposition of $Cu_2Te$ microdisks is proposed as the main growth mechanism. The formation process for the basal plane of $Cu_2Te$ microdisks is initialized by random deposition, as shown in Figure 4a for the first 5 min. The $Cu_2Te$ can preferentially deposit and form crystal seeds on the (2 1 1) and (3 4 6) surface, providing high-energy nucleation sites for the stack-on-stack deposition. As the reaction proceeds, the thin $Cu_2Te$ nanosheets gradually cover the basal surface of $Cu_2Te$ microdisks with good compactness and high density, thus forming the stack-on-stack morphology of $Cu_2Te$ microdisks.

In order to investigate the optical properties of single $Cu_2Te$ microdisk, an individual $Cu_2Te$ microdisk is optically pumped at room temperature by using an excitation laser with 532 nm wavelength and different optical powers. Figure 5a shows the schematic diagram of optical setup for micro-PL measurements. The pump laser is focused through a 50× objective (NA = 0.65) onto the single $Cu_2Te$ microdisk with a laser spot of ~1.0 μm diameter. The PL signals are collected by the same objective and measured by a monochromator. Figure 5b shows the pump power-dependent PL spectra of a single $Cu_2Te$ microdisk with the typical thickness of ~150 nm and edge length of ~4 μm. Upon laser excitation with different pump intensities in the range of 2.5–250 kW·cm$^{-2}$, a broad spontaneous emission band centered at 975 nm is observed, corresponding to a photon energy of 1.27 eV that is consistent with the bandgap of copper telluride. Interestingly, narrow peaks are observed in the visible range of 620–640 nm, which have not been observed before and maybe ascribed to defects of $Cu_2Te$. This is not a Raman band observed in narrow peaks in PL spectrum, since by considering the excitation wavelength at 532 nm and the observed peak at about 627 nm, a Raman shift of about 2800 cm$^{-1}$ is derived. The peak of $Cu_2Te$ is not in this range of wavenumbers. This emission band features various peaks at 622.5, 627.5, 628.8, 632 and 633.8 nm wavelengths. Some peaks show a nonlinear increase of the emission intensity by increasing the pump intensity, especially at the higher values of the excitation intensities. Where, for instance, the peak at 627.5 nm shows a different trend with respect to the linear one observed for the peak at 622.5 nm in Figure S2 of Supporting Information. The intensity of the broad emission at 975-nm increases slowly with the pump intensity increase and without the appearance of sharp peaks. Figure 5c shows the emission spectra centered at 627.5 and 628.8 nm for the $Cu_2Te$ microdisk with pump intensity increasing from 2.5 to 250 kW·cm$^{-2}$. Two broad spontaneous emission bands centered at 627.5 and 628.8 nm with a full width at half maximum (FWHM) of ~1.8 nm are obtained under the relatively low pump intensity (e.g., $P < 2.5$ kW·cm$^{-2}$). Quality factors ($Q$) extracted from the recorded line width ($\lambda_{FWHM}$ ~1.8 nm) as $Q = \lambda/\lambda_{FWHM}$ of about ~350, which is too low to compensate the cavity losses and giving the low ratio of signal to noise. When the pump intensity increased up to 250 kW·cm$^{-2}$, two sharp peaks centered at 627.5 and 628.8 nm with $\lambda_{FWHM}$ of ~0.4 and ~0.3 nm, thus giving quality factors $Q$ of ~1568 and ~2096, respectively. The high quality factors indicate that the resonant feedback is sufficient to compensate the cavity losses and exhibiting the high signal/noise ratio in the lasing spectra. Furthermore, Figure 5d shows the integrated emission intensity from single $Cu_2Te$ microdisk as a function of the pump intensity for two lasing wavelengths of 627.5 and 628.8 nm. The power threshold of $P_{th}$ ~125 kW·cm$^{-2}$ is estimated for dual-wavelength lasing.

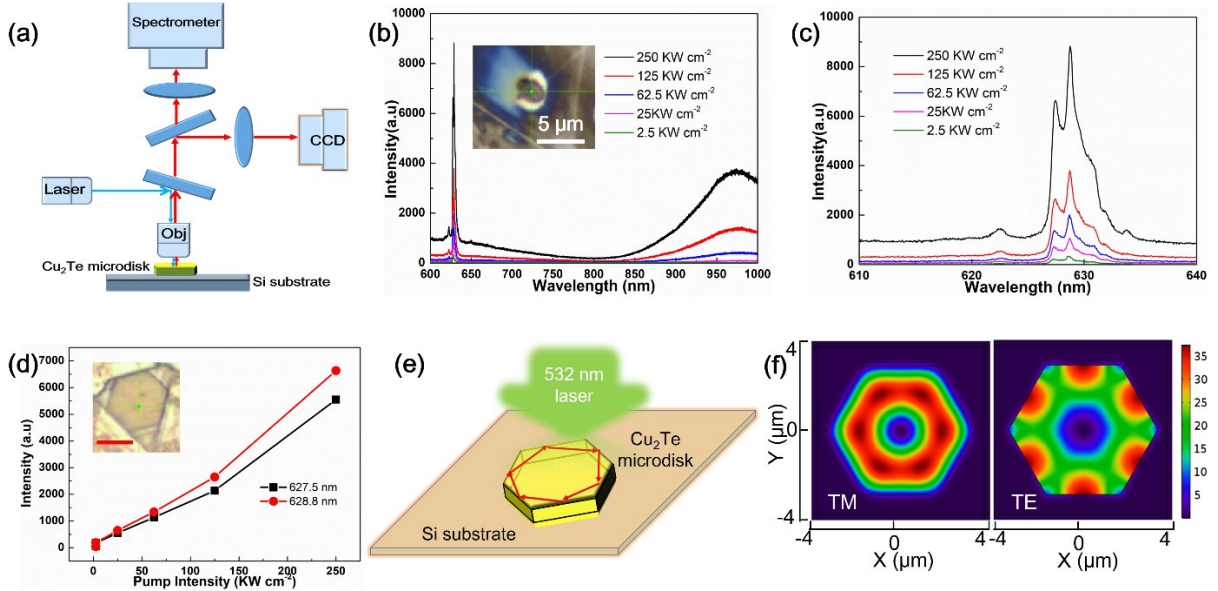

Figure 5. Lasing action of a single $Cu_2Te$ microdisk. (a) Schematic diagram of the micro-PL measurements. (b) Lasing spectra (covering the NIR waveband) of single $Cu_2Te$ microdisk with different pump intensity at room temperature. The inset shows an optical micrograph of the studied microdisk. (c) Detailed lasing spectra centered at 627.5 and 628.8 nm for the $Cu_2Te$ microdisk. (d) The integrated emission intensity as a function of the pump intensity for two lasing wavelengths of 627.5 and 628.8 nm. The inset shows the optical microscope picture of a hexagonal-shaped single $Cu_2Te$ microdisk. Scale bar: 4 µm. (e) Schematic diagram for the total internal reflection of PL signals excited by 532 nm laser in single $Cu_2Te$ microdisk. (f) Electromagnetic field distributions of the resonant cavity modes (TM, transverse magnetic mode; TE, transverse electric mode) for single $Cu_2Te$ microdisk with typical edge length of 4 µm.

$Cu_2Te$ hexagonal microdisks can constitute optical cavities for the confinement and amplification of light by multiple reflections at the edges of the microdisks, as schematized in Figure 5e. Various optical modes can be present, whose characteristic wavelengths ($\lambda_l$) and mode spacing ($\Delta\lambda$) can be estimated by a semi-classical approach for cavities having a size much larger than the wavelength of the light emitted by the active medium[37], obtaining as the following equation: $\Delta\lambda = \lambda_{l+1}^{-1} - \lambda_l^{-1} = (A \cdot n_i \cdot a)^{-1}$, where $\lambda_{l+1}$ and $\lambda_l$ are the wavelengths of two consecutive optical modes with mode index $l$, $n_i$ and $a$ are the refractive index and the length of the edge of the hexagonal microdisk, respectively, and $A$ is a constant that is equal to $3\sqrt{3}$ or 9 for modes with hexagonal or triangular periodic loops, respectively. By considering $n_{Cu2Te}$ = 2.1 (calculated from *Reference 13*) and $a = 4$ μm, a mode spacing in the range $1–2 \times 10^{-2}$ μm$^{-1}$ is calculated, a range of values that well matches with the measured spacing between the various peaks of the red emission of single microdisks (about $1.2 \times 10^{-2}$ μm$^{-1}$).

To better understand the modes supported by the Cu$_2$Te hexagonal microdisk and the optical field distribution of the resonant cavity modes, numerical simulations are performed. By considering the resonant wavelength of 627.5 nm as an example, and the corresponding effective refractive index value of Cu$_2$Te, $n_{Cu2Te} = 2.1$, only two eigenmodes of transverse magnetic (TM) and electric (TE) modes have an effective index higher than that of surrounding medium, thus can support the total internal reflection in the Cu$_2$Te microdisk cavity, as schematically shown in Figure 5e. It suggests that a good mode confinement in the Cu$_2$Te microdisk is achieved, thus leading to an in-plane emission. Figure 5f shows the simulation results on the electric field distributions inside the hexagonal microdisk (edge length ~4 μm) for TM and TE modes with effective index of 2.18 and 1.97, respectively. In these two scenarios, the optical fields are well confined inside the microdisk cavities and the total internal reflection occurred between the hexagonal facets/corners, resulting in the formation of the WGMs. It can be interpreted that WGM characteristics origin from the confinement of planar waveguide mode by hexagonal structure.

Furthermore, the natural excitation of Cu$_2$Te LSPR in the NIR could be used in optical sensing such as SERS detection of R6G molecules, since copper vacancies provide Cu$_2$Te with a composition-dependent LSPRs in NIR spectrum[5, 6]. Plasmonic nanoparticles (NPs) find versatile applications for detection of biomolecules due to their extremely high near-electric fields to increase the signals of biomolecules close to their surfaces. Though noble metals such as small gold and silver NPs can exhibit LSPRs in the NIR region by increasing the particle size, there still remain limited applicability of these particles in living organisms. Thus, Cu$_2$Te, with LSPR in the NIR range could pave a way to SERS detection of biomolecules, as a low-cost and abundant nonmetallic SERS substrates. A systematic series of experiments was conducted to characterize the SERS detection capability of R6G molecules by immersion of $Cu_2Te$ microdisks in solutions with different concentration of R6G, as schematized in Figure 6a. We firstly tested the difference of Raman signals when collected from various substrates including bare Si and $Cu_2Te$ substrate with the same concentration of R6G solution being $10^{-4}$ mol (M), as shown in Figure 6b. Raman signals at 1650, 1506, 1360 and 1185 $cm^{-1}$ (the Raman peaks of the R6G molecule) with strong signal-to-noise ratio are observed in the spectrum of the $Cu_2Te$ substrate, while only a few vibration modes with poor signal-to-noise ratio can be seen in the Raman spectrum of the bare Si substrate. These results indicate the significance of $Cu_2Te$ microdisks for the enhancement of Raman signal, since R6G molecules are adsorbed efficiently on the surface of nonmetallic $Cu_2Te$ microdisks due to the electronegativity of copper vacancies and their interaction between molecules and $Cu_2Te$. Moreover, $Cu_2Te$ microdisks surface could generate a strong electric field resulting from plasmon resonance in the NIR spectrum influencing the R6G molecules filed with a very strong enhancement. Furthermore, Figure 5f illustrates the formation of the WGMs through the distribution of electromagnetic fields, which can also enhance the SERS activity of $Cu_2Te$ microdisks, in addition to the electromagnetic field enhancement induced by LSPR[5]. Moreover, charge transfer between R6G molecules and $Cu_2Te$ microdisks could also contribute to SERS enhancement, since electronic transition rate between $Cu_2Te$ and R6G molecule is increased by modulating their electronic and surface chemical properties. To validate ultra-detection of biomolecules by $Cu_2Te$ microdisks, the synthesized $Cu_2Te$ microdisks were immersed into R6G solution with sequential dilution concentration from $1\times10^{-4}$ M to $1\times10^{-9}$ M for 20 min for quantitative molecule detection. As shown in Figure 6c, Raman signals decrease with the decrease of R6G concentration.

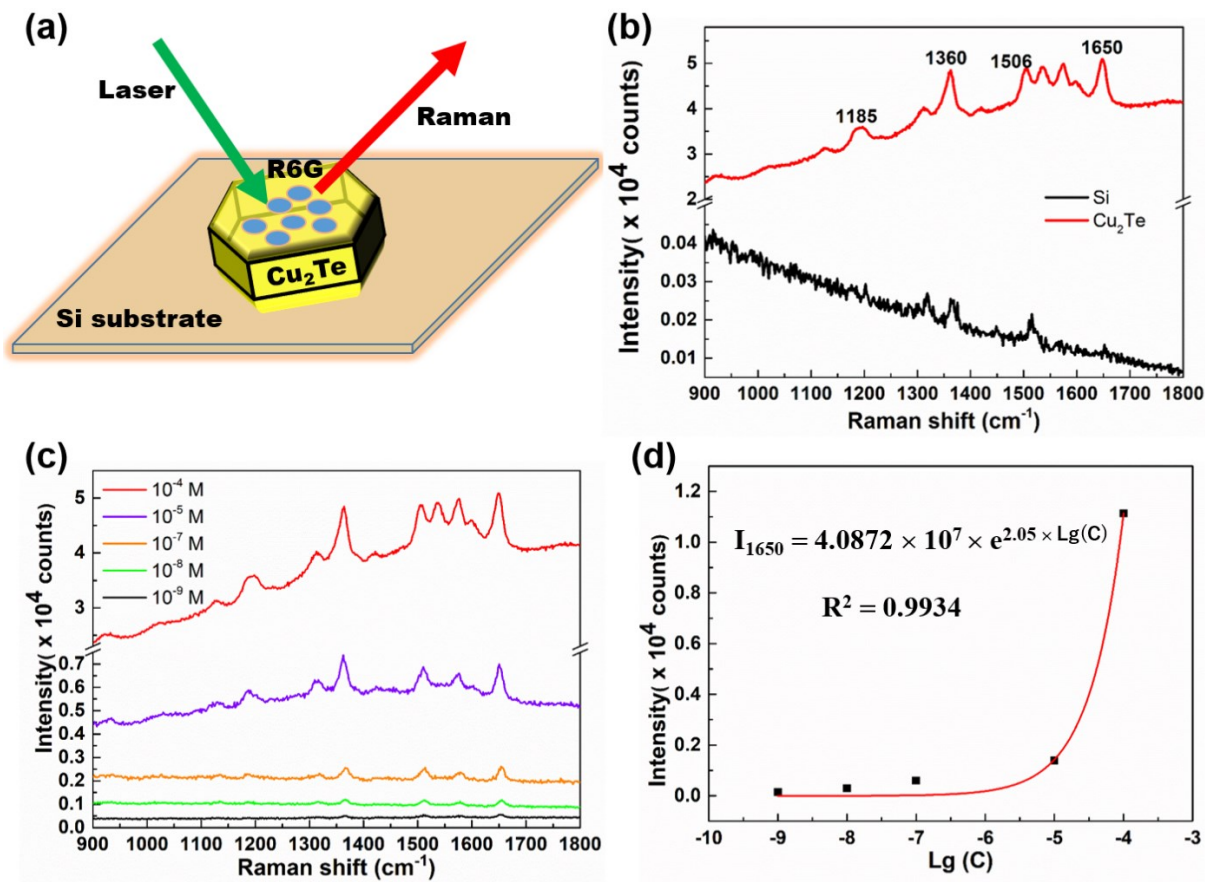

Figure 6. Performance analysis for SERS detection of R6G molecules based on $Cu_2Te$ microdisks. (a) Schematic of the Raman effect in the R6G/$Cu_2Te$ microdisks on Si substrate. (b) Raman spectra of R6G molecules deposited on Si (black curve) and $Cu_2Te$/Si (red curve) substrates, respectively. (c) Raman spectra of the R6G molecules deposited on $Cu_2Te$/Si substrates using solutions with different R6G concentrations from $1\times10^{-4}$ M to $1\times10^{-9}$ M. (d) Intensity of Raman signals at 1650 cm$^{-1}$ (after background subtraction) versus the logarithm of the R6G concentration [Lg(C)].

The intensity of the Raman peak at 1650 cm$^{-1}$ versus the logarithm of the R6G concentration [Lg(C)] is plotted in is plotted in Figure 6d, where the concentration levels for Raman detection of R6G molecules can be as low as $10^{-9}$ M. The Raman intensity ($I_{1650}$) versus the logarithm of the R6G concentration [Lg(C)] follows the index relationship of $I_{1650} = 4.0872 \times 10^7 \times e^{2.0519 \times Lg(C)}$ with good fitting regression coefficient ($R^2$) value of 0.9934, further indicating that $Cu_2Te$ can act as a SERS-active substrate for R6G detection. Though there exist a few SERS substrates with higher detection limit for R6G molecules, often requiring noble metals or complex structure such as Ag, Au NPs or grapheme, $Cu_2Te$ microdisks as nonmetallic substrate offer a versatile strategy with vacancy-mediated property to tune the SERS detection limit.

To investigate the influence of $Cu_2Te$ microdisks on the enhancement of Raman signal of R6G molecules, the analytical enhancement factors (*EF*) are calculated by dividing the Raman intensities of $Cu_2Te$ microdisks by those of R6G bulk crystals at 1650, 1506, 1360 and 1185 cm$^{-1}$, as shown in Figure 7a. The corresponding enhancement factor is calculated according to the equation[38]: $EF = (I_{SERS}/I_{bulk}) \times (N_{bulk}/N_{SERS})$, where *I* represent the Raman signals intensity of bands of symmetrical ring breathing mode for R6G molecules, *N* denoted as the number of R6G molecules adsorbed onto substrate, whereas $I_{SERS}$ is the Raman intensity of R6G on the $Cu_2Te$ substrate and $I_{bulk}$ is that of R6G under non-SERS condition, namely the Raman signal of R6G bulk crystals deposited on Si substrate here used. As an example, we choose Raman peak at 1650 cm$^{-1}$ for calculation, obtaining: *EF* ($Cu_2Te$) = $1.95 \times 10^5$ (details of the calculation are reported in *Supporting Information*). *EF*s of other Raman modes are lower than that of Raman peak at 1650 cm$^{-1}$, since the charge-transfer transition process is dominant at Raman peak at 1650 cm$^{-1}$. To identify the origin of this Raman enhancement effect from $Cu_2Te$ microdisk, valance band (VB) energy and conduction band (CB) energy[34] of $Cu_2Te$ with the energy band structure of R6G is plotted in Figure 7b. The highest occupied molecular orbital (HOMO) level of R6G molecule is −5.70 eV while the lowest unoccupied molecular orbital (LUMO) level is located at −3.40 eV[39]. Under 532 nm laser excitation, the electrons at the HOMO level of R6G molecule are easily excited to the LUMO, a so-called molecular resonance transfer[38,39] happens to improve the Raman enhancement effect since the energy gap of R6G at 2.30 eV is close to the excitation energy (2.33 eV) of 532 nm laser, resulting molecular resonance for SERS enhancement. Moreover, an electronic transition ($\mu_{CT}$) between the LUMO of R6G and the VB maximum (VBM) of $Cu_2Te$ is allowed due to the band-band transition energy of 0.8 eV, smaller than the excitation energy, and is expected to improve the Raman enhancement effect. Instead, the band-band transition energy between the HOMO of R6G and the CB minimum (CBM) of $Cu_2Te$ is 2.6 eV, which is larger than the excitation energy of 2.33 eV, hence such transition is unfavored and unlikely to contribute to the Raman enhancement effect.

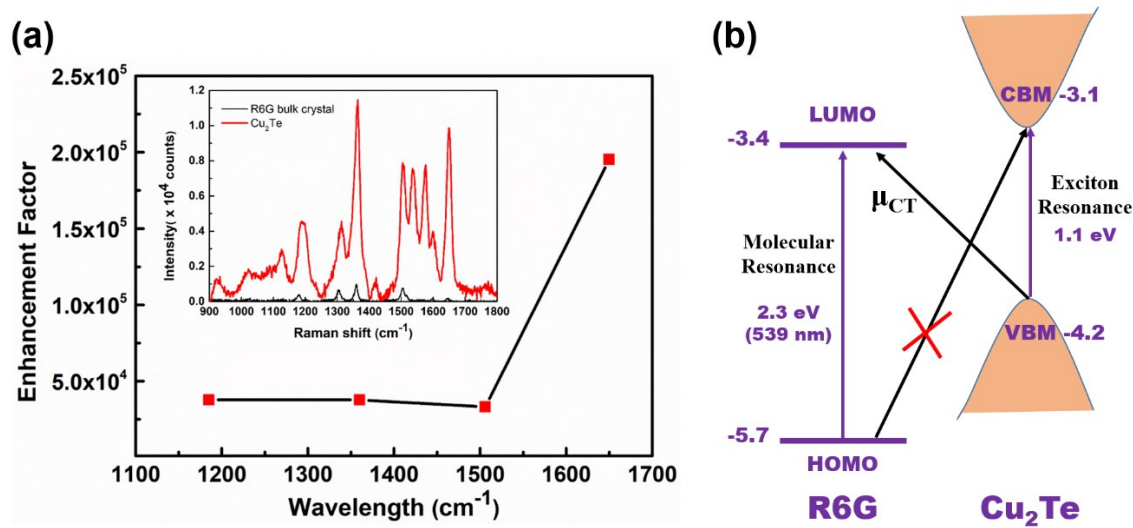

Figure 7. Raman enhancement mechanism of R6G molecules based on $Cu_2Te$ microdisks. (a) Enhancement factors of different vibrational modes of R6G molecules (molar concentration: $10^{-4}$ M) on $Cu_2Te$ microdisks. Inset shows the plot of Raman spectra from the $Cu_2Te$ substrate (red curve) and R6G bulk crystal (black) with background signal subtracted. (b) Energy level diagram of the R6G/$Cu_2Te$ heterostructure, describing the molecular resonance, exciton resonance and charge-transfer ($\mu_{CT}$) effect.

### 3. CONCLUSION

$Cu_2Te$ microdisks are successfully fabricated by a simple CVD method from Cu foam/sheet. The Cu foam/sheet acts as the Cu source with GaTe providing Te source for the favor growth of $Cu_2Te$, which yield the high surface area and large density of exposure resonator. The GaTe source tuned the $Cu_2Te$ to form the orthorhombic phase, which is demonstrated to be a good strategy for the phase control and then forming the hexagonal structure in $Cu_2Te$. In addition to the NIR PL emission, a PL band is newly observed in the visible range, which might be ascribed to the Cu deficiency in orthorhombic phase. Controlling the ratio of Cu versus Te may constitute a new degree of freedom to tune the bandgap of $Cu_2Te$ from near infrared to the visible range. Dual-wavelength lasing at 627.5 nm and 628.8 nm with quality factors of ~1568 and ~2096 were obtained from single $Cu_2Te$ hexagonal microdisk. The microdisk laser shows a power threshold of ~125 kW·cm$^{-2}$. Furthermore, the optical fields in microdisk cavity are interpreted by the total internal reflection supported WGM. The synthesized $Cu_2Te$ hexagonal microdisk used as the SERS substrate, were extraordinarily sensitive to R6G molecules with a nanomolar-level concentration and with an enhancement factor of ~$1.95 \times 10^5$. The copper vacancy in the $Cu_2Te$ lattice not only induced a large electronegativity on the surface causing R6G molecules easily absorbed onto substrate, but also contribute to an electromagnetic field enhancement resulting from surface plasmon resonances. Besides, the chemical mechanism of SERS, dipole-dipole interactions with R6G molecules and electronic transition between R6G and $Cu_2Te$, significantly improve the SERS performance and overcome the intrinsic limit of LSPR-based SERS. These effects synergistically improve the Raman enhancement in the $Cu_2Te$-based SERS substrate. Consequently, the semiconducting material of $Cu_2Te$ hexagonal microdisk is demonstrated to be a good candidate for red lasing and non-metallic SERS applications.

## 4. EXPERIMETNAL SECTION

**Preparation of $Cu_2Te$ microdisks.** $Cu_2Te$ microdisks were grown by a one-step CVD procedure. Firstly, an alumina boat containing GaTe (~20 mg, 99.99%, from Aladin) as source for $Cu_2Te$ growth was placed at the upstream in the furnace. A 10 mm × 10 mm copper foam or sheet (after anhydrous ethanol clean and sonication) was placed as growth substrate upside-down the alumina boat and ~3 cm away from the GaTe source in the furnace. The furnace was evacuated to a residual pressure of 0.1 mbar and then filled with Ar gas to remove the residual oxygen. The temperature of the furnace was increased up to 850°C from room temperature in 85 minutes and then held at this temperature for 30 minutes. During growth, 600 sccm (standard state cubic centimeter per minute) of Ar and 20 sccm of $H_2$ were used as transporting gas to flush the furnace. After 30 min, the furnace was cooled down to room temperature under Ar gas flow.

**Material Characterizations.** The morphology of the $Cu_2Te$ microdisks was characterized with a Zeiss SEM operating at an accelerating voltage of 5 kV. Chemical compositions of the hybrid $Cu_2Te$ structures were examined by the energy dispersive EDS in the Zeiss SEM equipped with a Bruker AXS Quantax system working at 20 keV. The crystallinity of the prepared $Cu_2Te$ microdisks was measured by X-ray diffraction (XRD, X′ Pert PRO MPD). The surface valence states of the elements in the $Cu_2Te$ microdisks were investigated by X-ray photoelectron spectroscopy (XPS, PHI-QUANTERA-2). TEM and HRTEM micrograph were taken with a Tecnai-G2-F30 field emission transmission electron microscope operating at an accelerating voltage of 200 kV.

**Raman and PL Measurements.** Raman measurements of the $Cu_2Te$ microdisks were performed with a laser confocal microscope (Jobin Yvon Horiba, XploRA) through a 50× objective (NA = 0.65) onto the active material with a laser spot of ~1.0 μm diameter. The $Cu_2Te$ microdisks were cast on the silicon substrate after ultrasonication. Raman measurements were performed under backscattering geometry with 532 nm laser excitation. The spectra were recorded from 20 to 2000 $cm^{-1}$ spectral range using an ultralow frequency filter, 1800 grooves/mm grating, and Peltier cooled charge coupled device (CCD) with a 150 s acquisition time. PL and lasing measurements were performed with 532 nm laser excitation under different excitation intensity.

**Numerical Simulation.** The optical field distributions were obtained by using a commercial electromagnetic software (COMSOL) with finite element (FE) solutions. Numerical simulations of microcavity optical modes were performed to study the optical feedback mechanism in $Cu_2Te$ microdisk laser deposited on silicon substrates. For simplicity, the effective index of refraction-planar waveguide model is introduced in 2D system, and the effective index rather than material index is used for the optical field distribution of cavity modes.

**SERS Measurements.** R6G solution in ethanol was prepared at different concentrations ($10^{-4}$ to $10^{-9}$ M) by a sequential dilution process. $Cu_2Te$ microdisks was rinsed several times with ethanol to remove the free molecules, then was immersed in the R6G solution for 30 min as preparation for the SERS substrate. The SERS signal was collected by confocal Raman spectroscopy (Jobin Yvon Horiba, XploRA) under a 532 nm excitation laser with laser power and acquisition time being 2.8 mW and 1 s, respectively.

## ACKNOWLEDGEMENTS

The work was supported by National Natural Science Foundation of China (Grant No. 11804120), The Professorial and Doctoral Scientific Research Foundation of Huizhou University (Grants No. 2020JB043) and The Program for Innovative Research Team of Guangdong Province & Huizhou University (IRTHZU). D.P. acknowledges the Italian Minister of University and Research PRIN 2017PHRM8X project ("3D-Phys"), and the PRA_2018_34 ("ANISE") project from the University of Pisa.

**REFERENCES**

(1) Farag, B. S.; Khodier, S. A. Direct and Indirect Transitions in Copper Telluride Thin Films. *Thin Solid Films*, **1991**, 201, 231.

(2) Ferizovic, D.; Muñoz, M. Optical, Electrical and Structural Properties of $Cu_2Te$ Thin Films Deposited by Magnetron Sputtering. *Thin Solid Films* **2011**, 519, 6115–6119.

(3) Zhou, J.; Wu, X.; Duda, A.; Teeter, G.; Demtsu, S. The Formation of Different Phases of $Cu_xTe$ and Their Effects on CdTe/CdS solar cells. *Thin Film Solids* **2007**, 515, 7364–7369.

(4) Han, C.; Li, Z.; Li, W.-j.; Chou, S.-l.; Dou, S.-x. Controlled Synthesis of Copper Telluride Nanostructures for Long-Cycling Anodes in Lithium Ion Batteries. *J. Mater. Chem. A* **2014**, 2, 11683−116690.

(5) Li, W.; Zamani, R.; Gil, P. R.; Pelaz, B.; Ibanez, M.; Cadavid, D.; Shavel, A.; Alvarez-puebla, R. A.; Parak, W. J.; Arbiol, J.; Cabot, A. CuTe Nanocrystals: Shape and Size Control, Plasmonic Properties, and Use as SERS Probes and Photothermal Agents. *J. Am. Chem. Soc.* **2013**, 135, 7098−7101.

(6) Kriegel, I.; Jiang, C.; Rodríguez-Fernández, J.; Schaller, R. D.; Talapin, D. V.; da Como, E.; Feldmann, J. Tuning the Excitonic and Plasmonic Properties of Copper Chalcogenide Nanocrystals. *J. Am. Chem. Soc.* **2012**, 134, 1583.

(7) Mukherjee, S.; Femi, O.E.; Chetty, R.; Chattopadhyay, K.; Suwas, S.; Mallik, R. C. Microstructure and Thermoelectric Properties of $Cu_2Te$-$Sb_2Te_3$ Pseudo-Binary System. *Appl. Surf. Sci.* **2018**, 449, 805-814.

(8) Yao, Y.; Zhang, B. P.; Pei, J.; Sun, Q.; Nie, G.; Zhang, W. Z.; Zhuo, Z. T.; Zhou, W. High Thermoelectric Figure of Merit Achieved in $Cu_2S_{1-x}Te_x$ Alloys Synthesized by Mechanical Alloying and Spark Plasma Sintering. *ACS Appl. Mater. Interfaces* **2018**, 10 (38), 32201−32211.

(9) Ghosh, A.; Mitra, M.; Banerjeeb, D.; Mondal, A. Facile Electrochemical Deposition of $Cu_7Te_4$ Thin Films with Visible-Light Driven Photocatalytic Activity and Thermoelectric Performance. *RSC Adv.* **2016**, 6, 22803-22811.

(10) Kumaravel, S.; Karthick, K.; Thiruvengetam, P.; Johny, J. M.; Sankar, S. S.; Kundu, S. Tuning Cu Overvoltage for a Copper–Telluride System in Electrocatalytic Water Reduction and Feasible Feedstock Conversion: A New Approach. *Inorg. Chem.* **2020**, 59 (15), 11129-11141.

(11) Wan, B.; Hu, C.; Zhou, W.; Liu, H.; Zhang, Y. Construction of Strong Alkaline Hydrothermal Environment for Synthesis of Copper Telluride Nanowires. *Solid State Sci.* **2011**, 13 (10), 1858-1864.

(12) Zhang, Y.; Qiao, Z.-P.; Chen, X.-M. Microwave-Assisted Elemental Direct Reaction Route to Nanocrystalline Copper Chalcogenides CuSe and $Cu_2Te$. *J. Mater. Chem.* **2002**, 12, 2747.

(13) Kriegel, I.; Rodríguez-Fernández, J.; Wisnet, A.; Zhang, H.; Waurisch, C.; Eychmüller, A.; Dubavik, A.; Govorov, A. O.; Feldmann, J. Shedding Light on Vacancy-Doped Copper Chalcogenides: Shape-Controlled Synthesis, Optical Properties, and Modeling of Copper Telluride Nanocrystals with Near-Infrared Plasmon Resonances. *ACS Nano* **2013**, 7, 4367.

(14) Li, H.; Brescia, R.; Povia, M.; Prato, M.; Bertoni, G.; Manna, L.; Moreels, I. Synthesis of Uniform Disk-Shaped Copper Telluride Nanocrystals and Cation Exchange to Cadmium Telluride Quantum Disks with Stable Red Emission. *J. Am. Chem. Soc.* **2013**, 135, 12270.

(15) Nowotny, H. Die Kristallstruktur von $Cu_2Te$. *Z. Metallkd.* **1946**, 37, 40.

(16) Blachnik, R.; Lasocka, M.; Walbrecht, U. The System Copper-Tellurium. *J. Solid State Chem.* **1983**, 48, 431.

(17) Nguyen, M. C.; Choi, J.-H.; Zhao, X.; Wang, Ho, C.-Z. K.-M.; Zhang, Z. New Layered Structures of Cuprous Chalcogenides as Thin Film Solar Cell Materials: $Cu_2Te$ and $Cu_2Se$. *Phys. Rev. Lett.* **2013**, 111, 165502.

(18) Qian, K.; Gao, L.; Li, H.; Zhang, S.; Yang, J.; Liu, C.; Wang, J.; Qian, T.; Ding, H.; Zhang, Y.; Lin, X.; Du, S.; Gao, H. Epitaxial Growth and Air-Stability of Monolayer $Cu_2Te$. *Chin. Phys. B* **2020**, 29, 018104.

(19) Miyatani, S.-y.; Mori, S.; Yanagihara, M. Phase Diagram and Electrical Properties of $Cu_{2-\delta}Te$. *J. Phys. Soc. Jpn*. **1979**, 47, 1152.

(20) Vouroutzis, N.; Manolikas, C. Phase Transformations and Discommensurations in Hexagonal Cuprous Telluride. *Phys. Status Solidi A* **1989**, 115, 399.

(21) Vouroutzis, N.; Manolikas, C. Phase Transformations in Cuprous Telluride. *Phys. Status Solidi A* **1989**, 111, 491.

(22) Pashinkin, A. S.; Fedorov, V. A. Phase Equilibria in the Cu−Te System. *Inorg. Mater.* **2003**, 39, 539.

(23) Da Silva, J. L. F.; Wei, S.-H.; Zhou, J.; Wu, X. Stability and Electronic Structures of $Cu_xTe$. *Appl. Phys. Lett.* **2007**, 91, 091902.

(24) Kashida, S.; Shimosaka, W.; Mori, M.; Yoshimura, D. Valence Band Photoemission Study of the Copper Chalcogenide Compounds, $Cu_2S$, $Cu_2Se$ and $Cu_2Te$. *J. Phys. Chem. Solids* **2003**, 64, 2357.

(25) Sridhar, K.; Chattopadhyay, K. Synthesis by Mechanical Alloying and Thermoelectric Properties of $Cu_2Te$. *J. Alloys Compd.* **1998**, 264, 293.

(26) He, Y.; Zhang, T.; Shi, X.; Wei, S.-H.; Chen, L. High Thermoelectric Performance in Copper Telluride. *NPG Asia Mater.* **2015**, 7, e210.

(27) Mansour, B.; Mukhtar, F.; Barakati, G. G. Electrical and Thermoelectric Properties of Copper Tellurides. *Phys. Status Solidi A* **1986**, 95, 703.

(28) Ballikaya, S.; Chi, H.; Salvador, J. R.; Uher, C. Thermoelectric Properties of Ag-Doped $Cu_2Se$ and $Cu_2Te$. *J. Mater. Chem. A* **2013**, 1, 12478.

(29) Zhao, K. P.; Liu, K.; Yue, Z. M.; Wang, Y. C.; Song, Q. F.; Li, J.;Guan, M. J.; Xu, Q.; Qiu, P. F.; Zhu, H.; Chen, L. D.; Shi, X. Are $Cu_2Te$-Based Compounds Excellent Thermoelectric Materials? *Adv.Mater.* **2019**, 31, 1903480.

(30) Han, C.; Bai, Y.; Sun, Q.; Zhang, S.; Li, Z.; Wang, L.; Dou, S. Ambient Aqueous Growth of $Cu_2Te$ Nanostructures with Excellent Electrocatalytic Activity Toward Sulfide Redox Shuttles. *Adv. Sci.* **2016**, 3, 1500350.

(31) Ding, X.; Liow, C. H.; Zhang, M.; Huang, R.; Li, C.; Shen, H.; Liu, M.; Zou, Y.; Gao, N.; Zhang, Z.; Li, Y.; Wang, Q.; Li, S.; Jiang, J. Surface Plasmon Resonance Enhanced Light Absorption and Photothermal Therapy in the Second Near-Infrared Window. *J. Am. Chem. Soc.* **2014**, 136, 15684.

(32) Chen, H.; Song, M.; Tang, J.; Hu, G.; Xu, S.; Guo, Z.; Li, N.; Cui, J.; Zhang, X.; Chen, X.; Wang, L. Ultrahigh 19F Loaded $Cu_{1.75}S$ Nanoprobes for Simultaneous 19F Magnetic Resonance Imaging and Photothermal Therapy. *ACS Nano* **2016**, 10, 1355.

(33) Lv, R.; Yang, P.; He, F.; Gai, S.; Yang, G.; Lin, J. Hollow Structured $Y_2O_3$:Yb/Er−$Cu_xS$ Nanospheres with Controllable Size for Simultaneous Chemo/Photothermal Therapy and Bioimaging. *Chem. Mater.* **2015**, 27, 483.

(34) Arciniegas, M. P.; Stasio, F. Di; Li, H.; Altamura, D.; De Trizio, L.; Prato, M.; Scarpellini, A.; Moreels, I.; Krahne, R.; Manna, L. Self Assembled Dense Colloidal $Cu_2Te$ Nanodisk Networks in P3HT Thin Films with Enhanced Photocurrent. *Adv. Funct. Mater.* **2016**, 26, 4535−4542.

(35) Salmón-Gamboa, J.U., Barajas-Aguilar, A.H., Ruiz-Ortega, L.I. et al. Vibrational and Electrical Properties of $Cu_{2–x}Te$ Films: Experimental Data and First Principle Calculations. *Sci. Rep.* **2018**, 8, 8093.

(36) Pandey, J.; Mukherjee, S.; Rawat, D.; Athar, S.; Rana, K. S.; Mallik, R. C.; Soni, A. Raman Spectroscopy Study of Phonon Liquid Electron Crystal in Copper Deficient Superionic Thermoelectric $Cu_{2–x}Te$. *ACS Appl. Energy Mater.* **2020**, 3, 2175−2181.

(37) Yang, Y.-D.; Tang, M.; Wang, F.-L.; Xiao, Z.-X.; Xiao, J.-L.; Huang, Y.-Z. Whispering-Gallery Mode Hexagonal Micro-/Nanocavity Lasers. *Photonics Res.* **2019**, 7, 594–607.

(38) Zheng, Z. H.; Cong, S.; Gong, W. B.; Xuan, J. N.; Li, G. H.; Lu, W. B.; Gengm, F. X.; Zhao, Z. G. Semiconductor SERS enhancement enabled by oxygen incorporation. *Nat. Commun.* **2017**, 8, 1993.

(39) Seo, J.; Lee, J.; Kim, Y.; Koo, D.; Lee, G.; Park, H. Ultrasensitive Plasmon-Free Surface Enhanced Raman Spectroscopy with Femtomolar Detection Limit from 2D van der Waals Heterostructure. *Nano Lett.* **2020**, 20, 1620−1630.

# Supporting Information

# Unusual Red Light Emission from Nonmetallic Cu$_2$Te Microdisk for Laser and SERS Applications

*Qiuguo Li,* [1] *Hao Rao,* [2] *Xinzhou Ma,* [3] *Haijuan Mei,* [1] *Zhengting Zhao,* [1] *Weiping Gong,\**[,1] *Andrea Camposeo* [4]*, Dario Pisignano* [4,5] *and Xianguang Yang\**[,2]

[1]Guangdong Provincial Key Laboratory of Electronic Functional Materials and Devices, Huizhou University, Huizhou 516001, Guangdong, China

[2]Institute of Nanophotonics, Jinan University, Guangzhou 511443, China

[3]School of Materials Science and Energy Engineering, Foshan University, Foshan, 528000, China

[4]NEST, Istituto Nanoscienze-CNR and Scuola Normale Superiore, Piazza S. Silvestro 12, I-56127 Pisa, Italy

[5]Dipartimento di Fisica, Università di Pisa, Largo B. Pontecorvo 3, I-56127 Pisa, Italy

To clearly show the hexagonal structure, Figure S1a and S1b show magnified SEM images of two areas highlighted with yellow (a) and red (b) dotted lines in Figure 1b, respectively, indicating that the structure of the grown $Cu_2Te$ microdisks is composed by stacked layers.

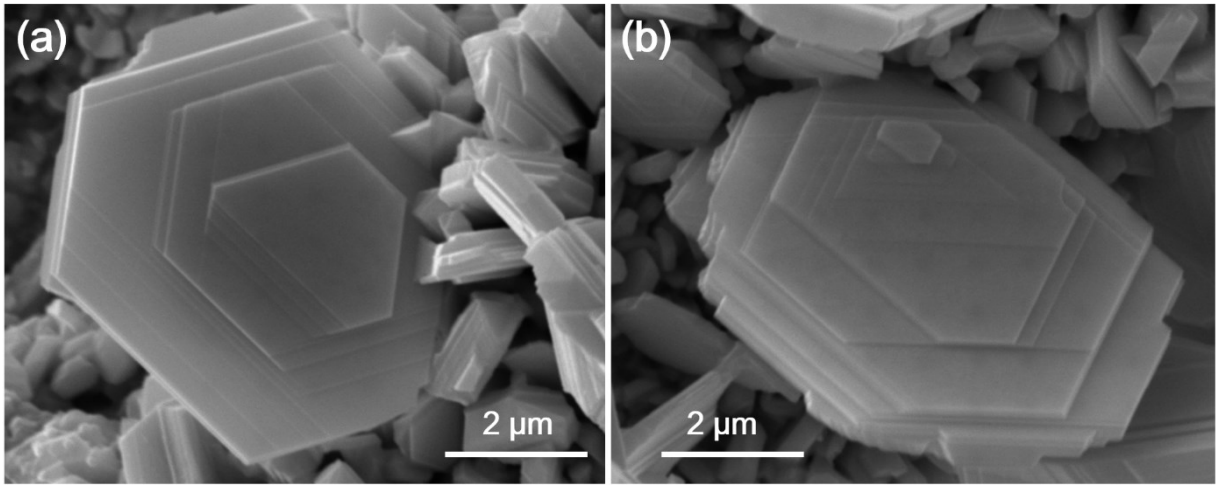

Figure S1. (a, b) SEM micrographs showing the hexagonal structure with stacking layers. The SEM micrographs are collected in areas highlighted by a yellow (a) and red (b) dotted lines in Figure 1b.

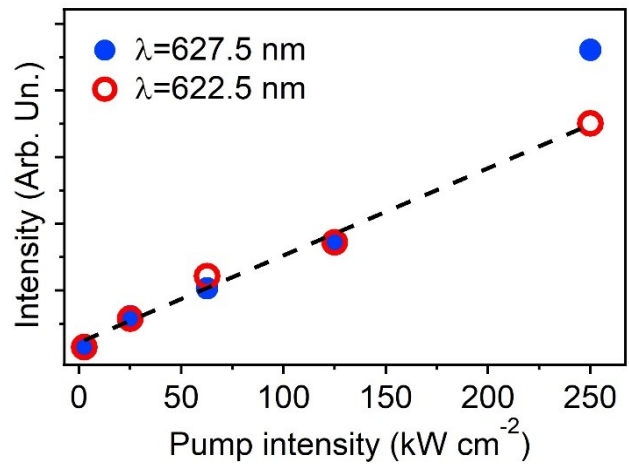

Figure S2. Emission intensity of the peaks at $\lambda$=622.5 nm (red empty circles) and $\lambda$=627.5 nm (blue full circles) *vs*. pump intensity. The dashed line is a linear fit to the data of the peak at $\lambda$=622.5 nm.

**Calculation of enhancement factor (*EF*)**

The *EF* was calculated according to the equations [1]:

$$EF = (I_{SERS}/I_{bulk}) \times (N_{bulk}/N_{SERS}) \qquad (1)$$

$$N_{SERS} = C \cdot V \cdot N_A \cdot A_{laser}/A_{sub} \qquad (2)$$

$$N_{bulk} = \rho \cdot h \cdot A_{laser} \cdot N_A/M_w \qquad (3)$$

where, $I_{SERS}$ is the Raman intensity of R6G on the SERS substrate of $Cu_2Te$ microdisk and $I_{bulk}$ is the Raman signal of R6G bulk crystals deposited on Si substrate (inset of Figure 7a). $N_{SERS}$ denotes the number of R6G molecules absorbed on the SERS substrate within the laser spot area $A_{laser}$ (~1 μm in diameter). $N_{SERS}$ can be calculated by equation (2) assuming the molecules are distributed uniformly on the SERS substrate. $C$ denotes the molar concentration of R6G solution ($10^{-4}$ M) while $V$ refers to the volume of droplet (~2 μL). The droplet was spread onto the SERS substrate with a circle of 4 mm in diameter after total evaporation, from which the effective area of the SERS substrate, $A_{Sub}$, can be obtained. $N_A$ ($6.023 \times 10^{23}$) is the Avogadro's constant. $N_{bulk}$ is the number of molecules within the laser spot area for R6G bulk crystals, which can be determined by using the equation (3). The confocal depth ($h$) of the laser beam into bulk crystal is 21 μm, and $M_w$ (479 g/mol) is the molecular weight and $\rho$ (1.15 g cm$^{-3}$) is the density of R6G bulk crystals, $N_{bulk}$ is calculated as $2.4 \times 10^{10}$.

As an example, we choose Raman peak at 1650 cm$^{-1}$ for calculation, taking all the parameters into consideration, the *EF* value can be estimated based on the following equations:

$N_{SERS} = C \cdot V \cdot N_A \cdot A_{laser}/A_{sub} = 10^{-4} \times 2 \times 10^{-6} \times 6.023 \times 10^{23} \times [\pi \times (0.5 \times 10^{-6})^2]/[\pi \times (2 \times 10^{-3})^2] = 7.53 \times 10^6$

$EF\ (Cu_2Te) = (I_{SERS}/I_{bulk}) \times (N_{bulk}/N_{SERS}) = (9800/160) \times (2.4 \times 10^{10}/7.53 \times 10^6) = 1.95 \times 10^5$

Therefore, we calculated the *EF* for R6G molecules on $Cu_2Te$ substrates at 1506, 1360 and 1185 cm$^{-1}$ (the Raman peaks of the R6G molecule) to be $0.33 \times 10^5$, $0.38 \times 10^5$, and $0.38 \times 10^5$, respectively.

**Reference**

[1]  Z. H. Zheng, S. Cong, W. B. Gong, J. N. Xuan, G. H. Li, W. B. Lu, F. X. Gengm, Z. G. Zhao, *Nat. Commun*. **2017**, *8*, 1993.